%% file: main.tex
\documentclass[10pt, conference, letterpaper, twocolumn]{IEEEtran}
 
\usepackage{amsmath,amssymb,amsfonts}
\usepackage{graphicx}
\usepackage{textcomp}
\usepackage[dvipsnames]{xcolor}

\usepackage{mathtools}  
\usepackage{amsmath}
\usepackage{amssymb}
\usepackage{tabulary}
\usepackage{booktabs}
\usepackage[inkscapearea=page]{svg}

\usepackage{paralist}

\usepackage{bbding}
\usepackage{pifont}
\usepackage{wasysym}
\usepackage{amssymb}
\usepackage{diagbox}

\usepackage[top=0.70in, left=0.58in, bottom=.96in, right=0.57in]{geometry}

\usepackage[font=small]{caption}
\usepackage{subcaption}
\usepackage{booktabs}
\usepackage{bbm}
\usepackage{algorithmic}
\usepackage{algorithm}
\usepackage[numbers, compress]{natbib}

\usepackage[hyphens]{url}

\usepackage{mathptmx}
\usepackage{amsmath}
\usepackage{amssymb}
\usepackage{amsthm}

\theoremstyle{definition}

\input{notation}

\begin{document}

\title{
    {
    \huge{
    Energy-Efficient Deadline-Aware Edge Computing: Bandit Learning with Partial Observations in Multi-Channel Systems 
    }}
}
{
    \author{
        \IEEEauthorblockN{Babak Badnava\IEEEauthorrefmark{1}, Keenan Roach\IEEEauthorrefmark{2}, Kenny Cheung\IEEEauthorrefmark{2}, Morteza Hashemi\IEEEauthorrefmark{1}, Ness B Shroff\IEEEauthorrefmark{3}}
        \IEEEauthorblockA{
        \IEEEauthorrefmark{1}Department of Electrical Engineering and Computer Science, University of Kansas\\
        \IEEEauthorrefmark{2}Universities Space Research Association (USRA) \\
        \IEEEauthorrefmark{3}Department of Electrical and Computer Engineering, The Ohio State University
        }
    }
}

\maketitle

\begin{abstract}
In this paper, we consider a task offloading problem in a multi-access edge computing (MEC) network, in which
 edge users can either use their local processing unit to compute their tasks or offload their tasks to a nearby edge server through multiple communication channels each with different characteristics.
The main objective is to maximize the energy efficiency of the edge users while meeting computing tasks deadlines.  
In the multi-user multi-channel offloading scenario, users are distributed with partial observations of the system states. 
We formulate this problem as a stochastic optimization problem and leverage \emph{contextual neural multi-armed bandit} models to develop an energy-efficient deadline-aware solution, dubbed E2DA. The proposed E2DA framework only relies on partial state information (i.e., computation task features) to make offloading decisions.  
Through extensive numerical analysis, we demonstrate that the E2DA algorithm can efficiently learn an offloading policy and achieve close-to-optimal performance in comparison with several baseline policies that optimize energy consumption and/or response time. 
Furthermore, we provide a comprehensive set of results on the MEC system performance for various applications such as augmented reality (AR) and virtual reality (VR).
\end{abstract}


\section{Introduction}\label{sec:Intro}
\vspace{-.1cm}
It is envisioned that the next generation wireless networks (5G-and-Beyond) will enable an unprecedented proliferation
of computationally-intensive applications, such as face recognition,
location-based AR/VR, and online 3D gaming~\cite{SeaGate-2019-State}. However, adoption of these resource-hungry applications will be negatively
affected by limited on-board computing and energy resources in edge devices. 
In fact, there is an ever-increasing demand for mobile computational power, while on-board resources remain constrained. 
In order to bridge this gap, multi-access edge computing (MEC)~\cite{Shi-2016-Edge} has been contemplated as a solution to supplement the computing capability of the
end-users. In contrast to the traditional cloud computing architectures, MEC leverages the radio access networks
(RANs) to boost the computing power in the proximity to the end-users, thereby enabling the users to offload
their computations to the MEC servers and achieve low-latency computations. 

However, 
practical MEC architectures face significant challenges including efficient resource management (computing, energy, communication), coordination among distributed users, and providing guaranteed quality of service (QoS) for latency-critical services. Moreover, it is critical that any MEC management framework fully incorporates the underlying 5G-and-Beyond RAN architecture, such as multi-connectivity and multi-channel communication between edge users and base stations (edge servers)~\cite{Yao-2022-Delay,Giordani-2016-Multi}. 
\begin{figure}[t]
    \centering
    \includegraphics[width=.9\linewidth, trim= 45mm 0mm 10mm 0mm, clip=true]{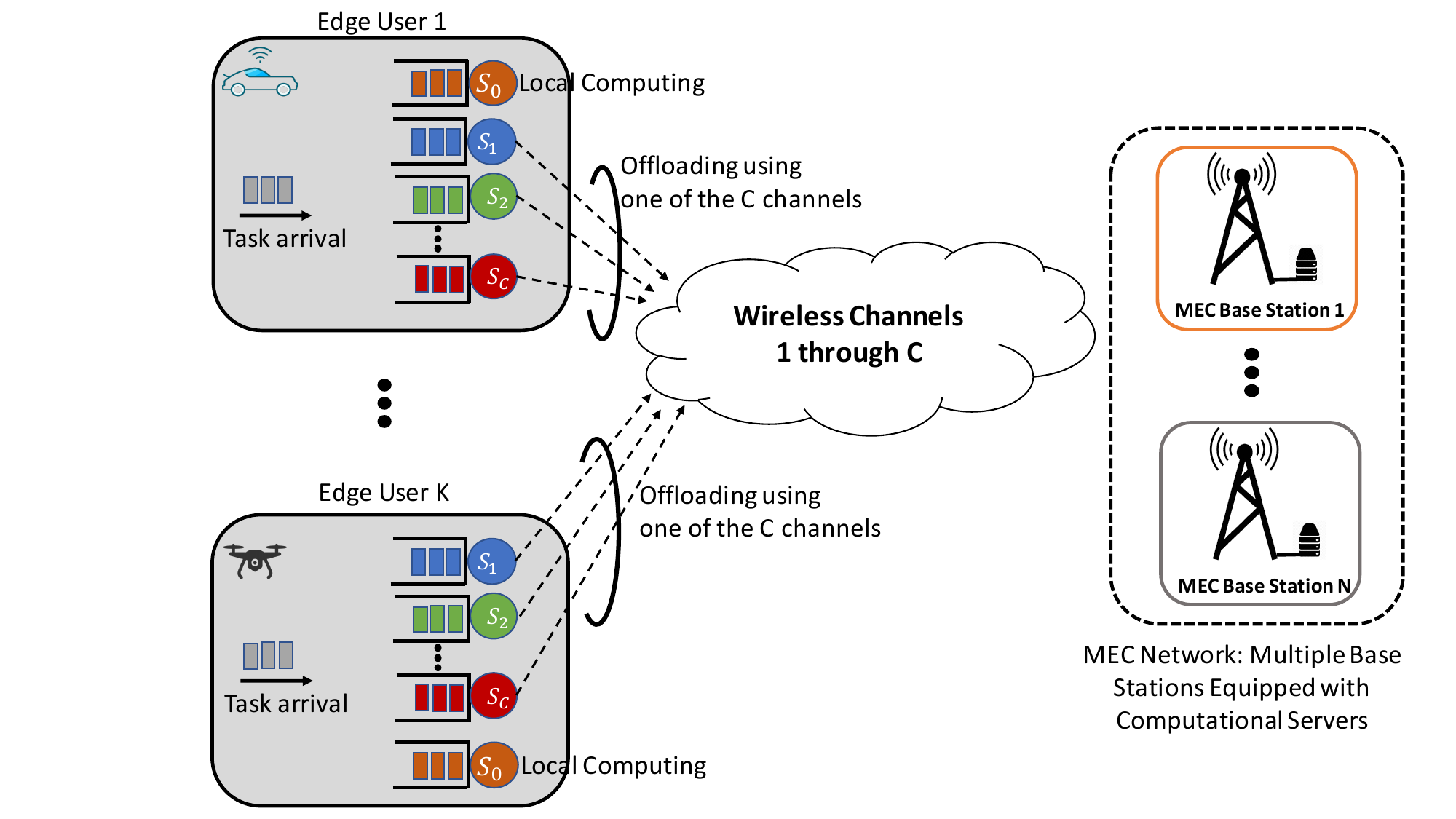}
    \caption{\small{Multi-User Multi-Channel MEC: Tasks arrive at the edge user randomly, and our E2DA agent decides, based on partial system information, to either compute each task locally or offload it to a nearby edge server through one of the $\nbChannels$ different wireless channels. }
    }
    \label{fig:sys-model}
    \vspace{-.25in}
\end{figure}
Within this context, there are three distinct sets of requirements that must be met: (i) \emph{network-imposed} constraints that determine the available communication links and their equality, as well as the number of edge servers, (ii) \emph{user-imposed} constraints that could be expressed in terms of available energy resources and local computing power, and (iii) \emph{task-imposed} constraints that dictate the required QoS (i.e., latency and deadline). To account for all these requirements, a decision-making agent is required to determine the optimal offloading policy.
The offloading policy could be initiated by either edge users or edge servers. User-initiated task offloading provides better-personalized services tailored to each user's individual preference, especially when edge servers are managed by different operators~\cite{Wang-2022-Decentralized}. 
However, distributed edge users only have access to \emph{partial observations} of the stochastic and dynamic system states (e.g., link rates, computational power of the edge server, total number of users in the system, etc.).
Hence, an efficient decision making policy should account for the stochastic nature of the system that affect latency and energy consumption, while being able to make offloading decisions in a partially observable environment.

A multitude of prior works have been devoted to investigate { and improve} various aspects of MEC systems. 
A group of work is dedicated to 
latency minimization in both partialy observable
\cite{Wang-2022-Decentralized,Ouyang-2019-Adaptive,Molina-2014-Joint}
and fully observable 
\cite{Wu-2021-EdgeCentric,Yang-2022-Optimal,Jia-2021-Learning,Huang-2019-Fine}
environments.
 Several works investigated the energy efficiency \cite{Liu-2020-Joint,Zhang-2018-Energy,Zhu-2018-Cooperative}
 and the effect of multi-connectivity technologies 
 \cite{Wu-2021-EdgeCentric,Sacco-2022-Self,Sacco-2021-Sustainable}
 on energy efficiency and latency in MEC systems.
However, to the best of our knowledge, no prior research fully integrates all the previously mentioned constraints in a  partially observable MEC setting. 
In this paper, we propose an energy efficient deadline-aware algorithm for task offloading in a decentralized multi-user multi-channel MEC setting, as depicted in Fig. \ref{fig:sys-model}.
First, we develop a complete system model that incorporates the network model, edge users characteristics, and computation task features.  
Next, we formulate a stochastic optimization problem and cast the problem of task offloading as a contextual multi-user multi-armed bandit (MAB) framework, where offloading to an edge server, using one of multiple available communication channels, is regarded as playing an arm. We develop a sequential energy-efficient delay-aware (E2DA) offloading scheme to balance the offloading exploration-exploitation trade-off. E2DA can learn the optimal offloading decision for each user, while only has access to its own computation task information such as task size, computational intensity, and deadline.
Our numerical results show that E2DA provides a promising service performance in comparison with three optimal baselines. In summary, the main contributions of this paper are as follows:
\begin{itemize}
    \item We introduce a partially observable decentralized multi-user multi-channel task offloading schema, where users of a MEC network can offload their tasks to a nearby edge server using different communication channels. We provide a queuing model of the proposed schema and formulate a stochastic optimization problem to find the optimal offloading decision and the best wireless channel.
    \item We propose a partially observable decentralized contextual neural multi-armed bandit algorithm to find the optimal task offloading policy that maximizes the energy efficiency.
    \item We develop a simulator based on standardized LTE implementation and perform extensive simulations to analyze the behavior of our proposed method. Our analysis shows that task computational intensity has little impact on edge users' energy consumption, while task size has a significant impact on energy consumption. Moreover, we show that task size has a significant impact on increasing task response time due to increased transmission and computation time.
\end{itemize}
The rest of this paper is organized as follows. 
In Section \ref{sec:system-model}, we present the system model. In section \ref{sec:solution}, we first formulate an energy-efficient deadline-aware task offloading optimization problem, and then presents our solution.
In Section \ref{sec:numerical-res}, we provide simulation results, and Section \ref{sec:conclusion} concludes the paper.

\vspace{-.1cm}
\section{System Model}\label{sec:system-model}
\vspace{-.1cm}
We consider a MEC network consisting of $\nbUAVs$ edge users and $\nbMECservers$ base stations, each of which is equipped with an edge server. 
As shown in Fig. \ref{fig:sys-model}, tasks arrive at the $k$th edge user according to an unknown random process (\eg Poison process). Then the scheduler (\ie E2DA agent) decides if the tasks should be computed locally or offloaded to the edge server.
Edge users can communicate with the edge servers through $\nbChannels$ different channels (\ie using carrier aggregation  or multiple radio access technology (RAT)) to offload their tasks. We assume that each edge user is associated with one base station at a time.


On the MEC side, edge servers allocate a virtual machine (VM) to each edge user. We assume that edge servers allocate the same amount of computational resources (\ie one CPU core with frequency $\mecCPUFreq$) to all the VMs, and tasks get service in a first-in-first-out (FIFO) manner.
After the task computation is done, the edge server sends back the task result to the corresponding edge user. 
In this section, we elucidate how we model tasks, their arrival at the edge users, the communication channel used to send tasks to edge servers, the computation model on both edge users and edge servers, and the energy consumption model for both communication and computation.




\subsection{Computation Task Model and Response Time}
\vspace{-.1cm}

\textbf{Task Model:} Tasks arrive at the edge users according to an unknown random process.
Each of these tasks is characterized by four different features: 
\begin{inparaenum}[(i)]
    \item task arrival time (in time), denoted by $\taskArrival$;
    \item task size/length (in bits),  denoted by $\taskSize$;
    \item task computational intensity (in CPU cycles per bit), denoted by $\taskIntencity$; and 
    \item task deadline (in seconds), denoted by $\taskDeadline$. 
\end{inparaenum}

As shown in Fig. \ref{fig:sys-model}, upon arrival, each task follows one of the two routes: it either gets computed locally or offloaded to the edge server for computation using one of the $\nbChannels$ communication channels. 
For example, in the case of an autonomous car as an edge user, object detection is a computationally intensive task that must be completed within some stringent deadline. In this scenario, the E2DA agent decides whether the object detection task should be computed locally or offloaded to an edge server.


In local computation, the scheduler adds the task to the processing queue at the edge user, and the task waits in the queue until the CPU picks it up for computation. Here, we assume that the CPU processes tasks in a FIFO manner. After the CPU finishes computing the task, the result of computation is ready, and the edge user can use that result, which is the point that the task exits the system. In the example of autonomous cars, the generated result is the image in which all cars and objects are detected.
On the other hand, when the scheduler decides to offload the task, it adds the task to the transmission queue that corresponds to the $c$'th communication channel. 
Then, on the MEC side, tasks are computed by the edge processing unit in a FIFO manner.
After computing a task, a result with the size of $\resultSize$ bits is generated to be sent back to the corresponding edge user over the same wireless channel used for offloading.

\textbf{Task Response Time.} To accurately model task response time, we note that each task could experience several types of delays, namely: (1) execution time, (2) transmission/reception time, and (3) queue waiting time. In the following, we calculate each of these delays separately.  

\noindent 
\emph{(1) Task execution time.}
From \cite{Badnava-2021-Spectrum}, the amount of time it takes a CPU with an operating frequency of $\cpuFreq$ (in cycle per second) to compute a task with size $\taskSize$ (in bits) and computational intensity $\taskIntencity$ (in cycles per bit) is calculated as:
\begin{align}\label{eq:cpu-exec-model}
    \taskExecution (\cpuFreq, \taskSize, \taskIntencity) &=  \frac{\taskSize \taskIntencity}{\cpuFreq} \quad [Seconds].
\end{align}
Hence, given CPU frequency of the allocated VM (\ie $\mecCPUFreq$) and edge user $\uavCPUFreq$, the execution time for offloaded tasks is calculated as $\taskExecution^{\mathit{offload}} = \frac{\taskSize \taskIntencity}{\mecCPUFreq}$, and the execution time for locally computed tasks is computed as $\taskExecution^{\mathit{local}} = \frac{\taskSize \taskIntencity}{\uavCPUFreq}$.

\noindent 
\emph{(2) Task transmission/reception time.}
We consider $\nbChannels$ different communication channels between the $k$-th edge user and the $n$-th base station. We denote $\ulChannelRate^c$ and $\dlChannelRate^c$ for the uplink and downlink transmission rates over the $c$-th wireless channel, respectively. Hence, task transmission time and result reception time are calculated using $\resultTxTime^c = \frac{\resultSize}{\dlChannelRate^c}$ and 
$\taskTxTime^c = \frac{\taskSize}{\ulChannelRate^c}$, respectively.

\noindent 
\emph{(3) Task queue waiting time.}
Each task goes through a different set of queues. 
A locally computed task waits for $\taskWaitTime_{1}$ seconds in the local processing queue. On the other hand, an offloaded task waits for $\taskWaitTime_{2}^c$ seconds in the transmission queue of the $c$-th wireless channel to be sent over the edge server, then waits for $\taskWaitTime_{3}$ seconds in the edge processing unit, and finally waits for $\taskWaitTime_{4}^c$ seconds in the transmission queue for its result to be sent back.
Given the three sources of delay, locally computed tasks only wait in the local processing queue to get service. Thus, we have:
\begin{equation}\label{eq:task-response-local}
    \begin{split}
        \taskResponse^\mathit{local} &= \taskWaitTime_{1} +  \taskExecution^{\mathit{local}} \quad [Seconds]. 
    \end{split}
\end{equation}
On the other hand, offloaded tasks  wait in two transmission queues to be transmitted over the air and the edge processing queue to get served. Thus, the response time is:
\begin{equation}\label{eq:task-response-offload}
    \begin{split}
        \taskResponse^\mathit{offload} = \taskWaitTime_{2}^c + \taskTxTime^c 
        + \taskWaitTime_{3} + \taskExecution^{\mathit{offload}} + \taskWaitTime_{4}^c + \resultTxTime^c \quad [Seconds].
    \end{split}
\end{equation}
\subsection{Energy Consumption Model}
The edge user's battery is depleted by three different components, (i) its CPU for local computation, (ii) its wireless transmitter for offloading tasks to the edge server, and (iii) its receiver for receiving results of computations. We model energy consumption of each of these components separately.

The main source of task execution energy consumption on edge devices is the CPU dynamic power usage~\cite{Zhang-2013-Energy}.
We adopt the CPU energy consumption model used in \cite{Zhang-2013-Energy} such that the energy consumption by a CPU with frequency $\uavCPUFreq$ to compute a task is 
$
\cpuEnergy = \kappa \taskSize \taskIntencity \uavCPUFreq^2,
$
where $\kappa$ is the CPU capacitance factor, $\taskSize$ is task size and $\taskIntencity$ is task computational intensity.

The power consumption for transmission and reception of a task over a wireless channel varies based on the used wireless technology (\eg Wi-Fi, LTE/5G)~\cite{Dusza-2013-CoPoMo}. 
We use a general model in which the transmitter energy consumption is given by:
\begin{align}\label{eq:tx-energy-consum}
    \txEnergy^c = \taskTxTime^c \ulPower^c \quad [Joule],
\end{align}
where $\taskTxTime^c$ denotes the task transmission time over the $c$-th wireless channel, and $\ulPower^c$ is the power allocation profile.
Similarly, the receiver energy consumption is:
\begin{align}\label{eq:rx-energy-consum}
    \rxEnergy^c = \resultTxTime^c \dlPower^c \quad [Joule].
\end{align}
where $\resultTxTime^c$ denotes the task result reception time over the $c$-th wireless channel, and $\dlPower^c$ is the reception power consumption profile.
Hence, the total energy consumed by the edge user is:
\begin{align}\label{eq:total-energy}
    \totalEnergy^c = \txEnergy^c + \cpuEnergy 
    + \rxEnergy^c \quad [Joule]. 
\end{align}
Note that $\txEnergy^c$ and $\rxEnergy^c$ are zero for local computation.   

\section{Energy-Efficient Deadline-Aware Offloading }\label{sec:solution}

Our goal is to design a policy that achieves the maximum energy efficiency (\ie  
$\taskResponse \totalEnergy^c / \taskSize$) while
ensuring that tasks deadlines are always met. This is critical for latency-sensitive applications such as autonomous cars. 
To this end, there are two decision variables. The first one is $u$, which corresponds to the offloading decision, and the second one is $c$, which corresponds to the index of the channel that will be used in case of offloading. Here, we formulate the following stochastic optimization problem:
\begin{equation}
\label{eq:optimization-problem}
\textbf{E2DA-Opt:} \begin{cases}
\mathop{\mathrm{max}}\limits_{u, c} & \expected{
            \frac{
                \taskSize
            }{
                \taskResponse \totalEnergy^c 
            }
        } \quad [(Bits/Second)/Joule] \\
        \text{subject to:} &  \taskResponse \leq \taskDeadline \\
        & \text{Eqs.} \; \ref{eq:task-response-local}, \ref{eq:task-response-offload}, \ref{eq:total-energy},
\end{cases}
\end{equation}
where $\taskSize$ is the task size, $\taskResponse$ denotes the task response time, $\totalEnergy^c$ is the task energy consumption, and $\taskDeadline$ denotes the task deadline. 
The first constraint makes sure that each task meets its deadline.
The rest of the constraints correspond to the system model as explained in section \ref{sec:system-model}.
This stochastic optimization problem is difficult and challenging to solve every time a new task arrives. 
One main challenge is that some task features (\eg transmission energy consumption, transmission time, waiting time, etc.) are not observable at the decision-making stage. 
Moreover, the environment is uncertain and stochastic due to time-varying wireless communication channel conditions. 
Thus, a robust offloading policy needs not only to capture the uncertainty of the environment but also makes its decision based on the partially observable state information at the offloading stage.

\textbf{Contextual Multi-Armed Bandit (CMAB) Approaches:}
CMAB solutions belong to the general class of online learning approaches by which a learner converges to the optimal solution through repeated interactions with an environment.  In each round, the learner is presented with a set of actions, each of which is associated with a multi-dimensional feature vector (\ie contextual information). After choosing an action (\ie playing an arm), the learner will receive a stochastic reward generated from some unknown distribution conditioned on the action’s feature vector. 
During the learning process, the learner tries to balance the trade-off between exploration and exploitation, and its goal is to maximize the expected cumulative rewards over a finite number of trials. 
CMAB algorithms have been applied to many real-world applications, such as video streaming quality of experience improvement~\cite{Badnava-2022-QOE} and recommendation systems~\cite{Tang-2015-Personalized}.
\citeauthor{Zhou-2020-Neural} in \cite{Zhou-2020-Neural} provide a comprehensive comparison of the regret growth for different synthetic datasets. To handle the trade-off between the implementation simplicity and regret bound, we leverage Neural $\epsilon$-Greedy algorithm, which uses a neural network to model the expected reward given the contextual information. 
This, in turn, will enable us to capture the non-linearity of the reward function. 



\textbf{Energy-Efficient Deadline-Aware CMAB:}
\begin{algorithm}[t]
\caption{Energy-Efficient Deadline-Aware CMAB}
\label{alg:training-alg}
\textbf{Inputs:} \\
\hspace*{\algorithmicindent} Neural network $Q$ initialized by $\theta_0$;
Task replay buffer $\mathcal{B}$; \\
\textbf{Algorithm:}
\begin{algorithmic}[1]
    \FOR{epoch $e=1$ to $\mathcal{E}$}
        \FOR{each user $k$ in $\nbUAVs$}
            \FOR{each task upon arrival}
                \STATE Observe contextual info. $x = \left[\taskSize, \taskIntencity, \taskDeadline \right]$
                \STATE $a = \mathit{\epsilon\textendash Greedy}(Q(x \textemdash \theta))$
                \IF{$a = 0 \; (local\; computing)$}
                    \STATE Compute $\taskResponse$ using Eq. \ref{eq:task-response-local}
                    \STATE $\totalEnergy = \cpuEnergy$
                \ELSE
                    \STATE $c=a$
                    \STATE Compute $\taskResponse$, $\txEnergy^c$, $\rxEnergy^c$ using Eq. \ref{eq:task-response-offload}, \ref{eq:tx-energy-consum}, \ref{eq:rx-energy-consum}.
                    \STATE $\totalEnergy = \txEnergy^c + \rxEnergy^c$
                \ENDIF
                \STATE $r = \indicator_{\taskResponse \leq \taskDeadline} ( \frac{\taskSize}{\taskResponse\totalEnergy} + \lambda) - \lambda$
                \STATE Add tuple $(x, a, r)$ to replay buffer $\mathcal{B}$
                \STATE Let $\theta = TrainNN(\mathcal{B}, \theta_0)$
            \ENDFOR
        \ENDFOR
    \ENDFOR
\end{algorithmic}
\end{algorithm}
Every CMAB problem is defined using a set of contextual information $X$ and a set of actions $A$. Here, the contextual information set is a multi-dimensional vector consisting of task information.
The task information, which is available during the decision-making stage, includes task size $\taskSize$, task computational intensity $\taskIntencity$, and task deadline $\taskDeadline$.
The action set $A$ is, however, a discrete set with $\nbChannels + 1$ possibilities (\ie one possibility for local computation, and one for each of the $\nbChannels$ wireless communication channels). We designate $a=0$ to local computation, and $a \in \{1, 2, ..., \nbChannels\}$ corresponding to the channel chosen for the offloading process. 

In order to solve the introduced energy efficient deadline-aware optimization problem, it is important to note that the objective function in Eq. \ref{eq:optimization-problem} and the objective of bandit problems are analogous to each other in the sense that the objective in both is to maximize the expected value of a term; $\frac{ \taskSize }{ \taskResponse \totalEnergy^c }$ in Eq. \ref{eq:optimization-problem} and reward in bandit problems.
However, bandit problems are designed for unconstrained problems, {while Eq. \ref{eq:optimization-problem} contains some equality and inequality constraints. Equality constraints (\ie Eqs. \ref{eq:task-response-local}, \ref{eq:task-response-offload}, and \ref{eq:total-energy}) are part of the system model (\ie MEC environment) and are already incorporated in the reward function, however the only ineqaulity constraint (\ie $\taskResponse \leq \taskDeadline$) is not included in the objective function. }
Thus, we need to incorporate the first inequality constraint, in Eq. \ref{eq:optimization-problem}, in the reward function in a way that the learner gets penalized whenever the decision it makes leads to not meeting the deadline.
Hence, the following reward function is considered:
$$
    r = \indicator_{\taskResponse \leq \taskDeadline} ( \frac{\taskSize}{\taskResponse\totalEnergy^c} + \lambda) - \lambda,
$$
where $\lambda$ is the penalization factor for tasks that have not met their deadlines, and $\indicator_{\taskResponse \leq \taskDeadline}$ is an indicator function that returns one only when the task meets its deadline.
Therefore, if a task would meet its deadline, the reward function gives away a reward equal to $\taskSize / \taskResponse\totalEnergy^c$ for the task in hand, and $-\lambda$ otherwise. 
Note that $\taskSize$ (\ie task size) for a given task is constant, however, both $\taskResponse$ (\ie task response time) and $\totalEnergy^c$ (\ie task energy consumption) are stochastic values influenced by wireless channel condition and the number of users using that channel.

The condition of the wireless channel depends on several factors such as user mobility, propagation environment, distances, etc.~\cite{TS36777}.
These factors introduce uncertainties in the decision-making process. 
The goal of the  E2DA policy is to capture these uncertainties, thereby achieving a policy that is able to make a decision in the presence of these uncertainties.
Furthermore,  it is important for the E2DA policy to consider the allocation of communication resources, as wireless channels may be utilized by multiple users simultaneously, which leads to a reduced reward per user.
Overall, Algorithm \ref{alg:training-alg} presents the complete training process of E2DA that implements the  $\epsilon$-Greedy CMAB algorithm based on the defined contextual information and reward function. 

\section{Numerical Results}\label{sec:numerical-res}
\vspace{-.1cm}

\subsection{Simulation Setting, Training, and Baseline Algorithms}
\vspace{-.1cm}
\textbf{{Communication and Computation Models: }} 
To create a set of computing tasks for E2DA training, we designed a simulation framework for MEC using the ns-3 network simulator. This discrete event simulator incorporates several communication technologies, such as LTE, 5G, and Wi-Fi.
While MEC is being standardized to be included in 5G-and-Beyond networks \cite{ETSI-2018-MEC}, 
we use the ns-3 LTE module that is more developed and stable compared with the ns-3 5G module \cite{Katerina-2022-Calibration}.
The edge users offload their tasks to the nearby base station using one of $\nbChannels = 3$ channels.
Each of these component carriers operates on a different frequency (\ie $700$ MHz, $1500$ MHz, and $2600$ MHz).



Computation tasks arrive at the $k$-th edge user according to a Poisson process with the mean arrival rate of $\lambda_k = 40$ tasks per second. 
The task size is set according to a uniform distribution with min $10$ bits per task and max $75$ Kbit per task; The task intensity is set according to a uniform distribution with min of $10$ cycles per bit and max of $1000$ cycles per bit; and the task deadline follows a uniform distribution with min of $10 ms$ and max of $18 ms$.

\begin{figure*}[t]
    \centering
    \begin{subfigure}{.24\textwidth}
        \includegraphics[width=\linewidth, trim= 0mm 20mm 0mm 10mm, clip=true]{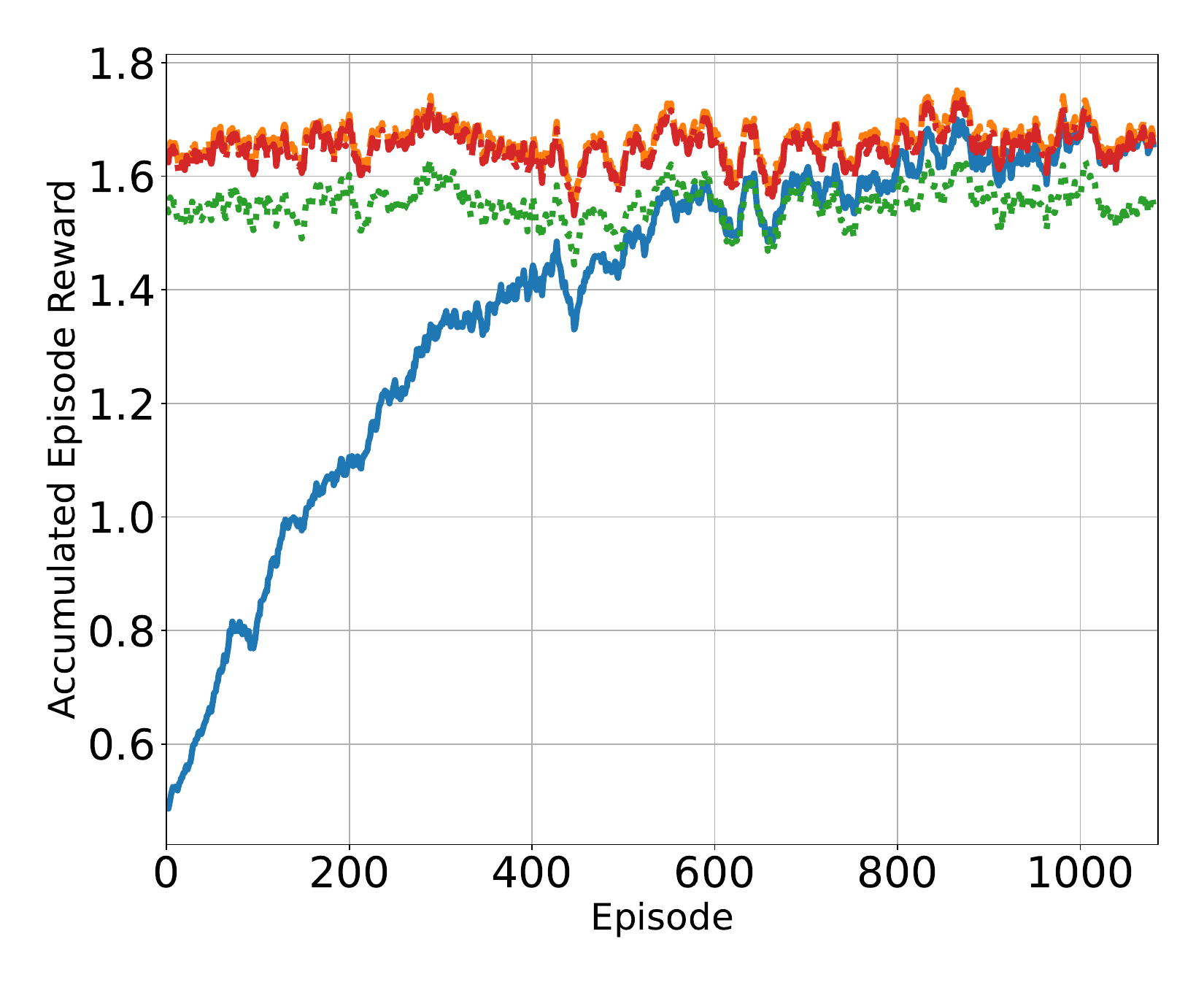}
        \caption{Learning Curve
        }
        \label{fig:mab-accumulated-reward}
    \end{subfigure}
    \hfill
    \centering
    \begin{subfigure}{.24\textwidth}
        \centering
        \includegraphics[width=\linewidth, trim= 0mm 20mm 0mm 10mm, clip=true]{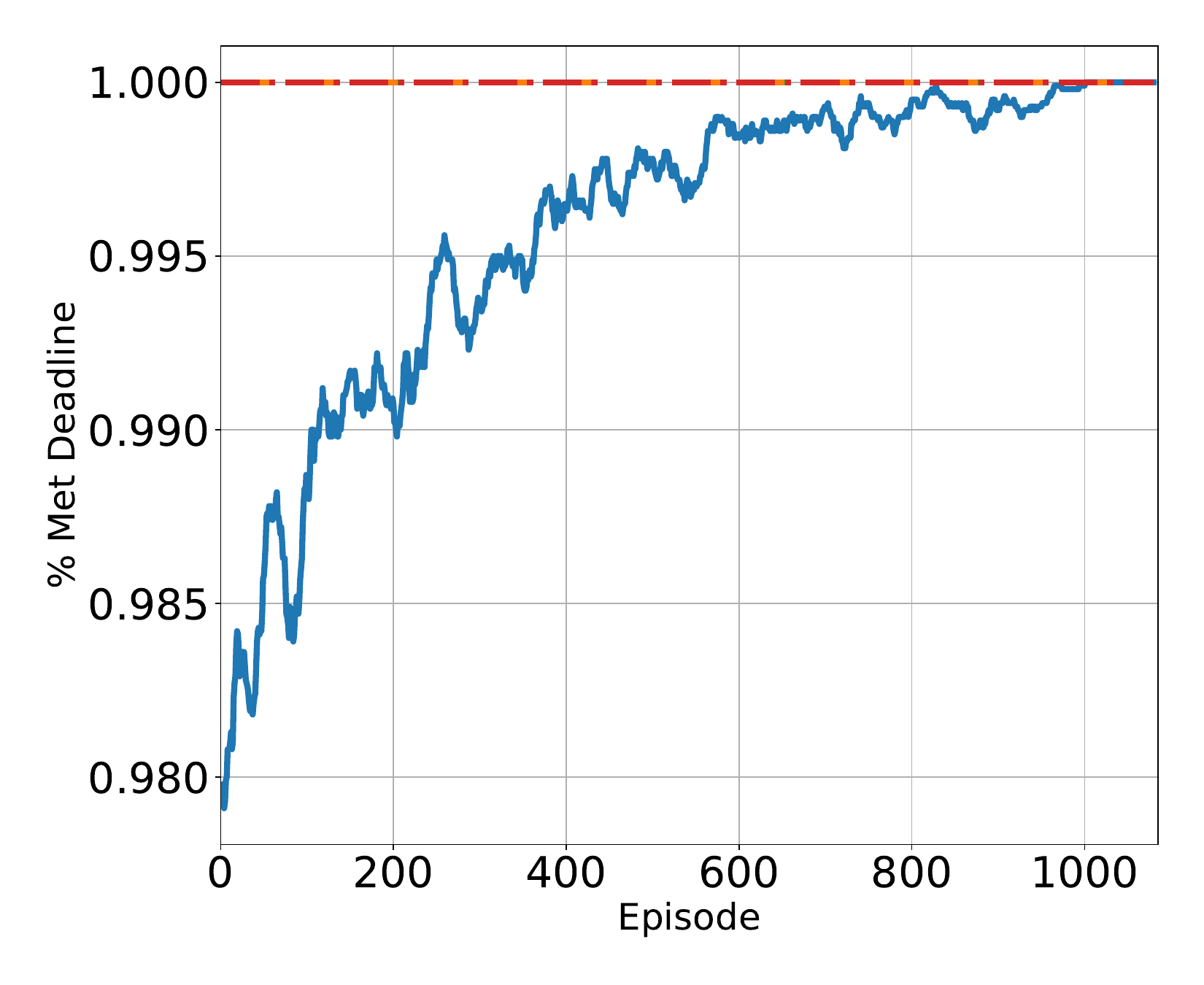}
        \caption{Percentage Met their Deadlines}
        \label{fig:mab-performance-analysis-a}
    \end{subfigure}
    \hfill
    \begin{subfigure}{.24\textwidth}
        \centering
        \includegraphics[width=\linewidth, trim= 0mm 20mm 0mm 10mm, clip=true]{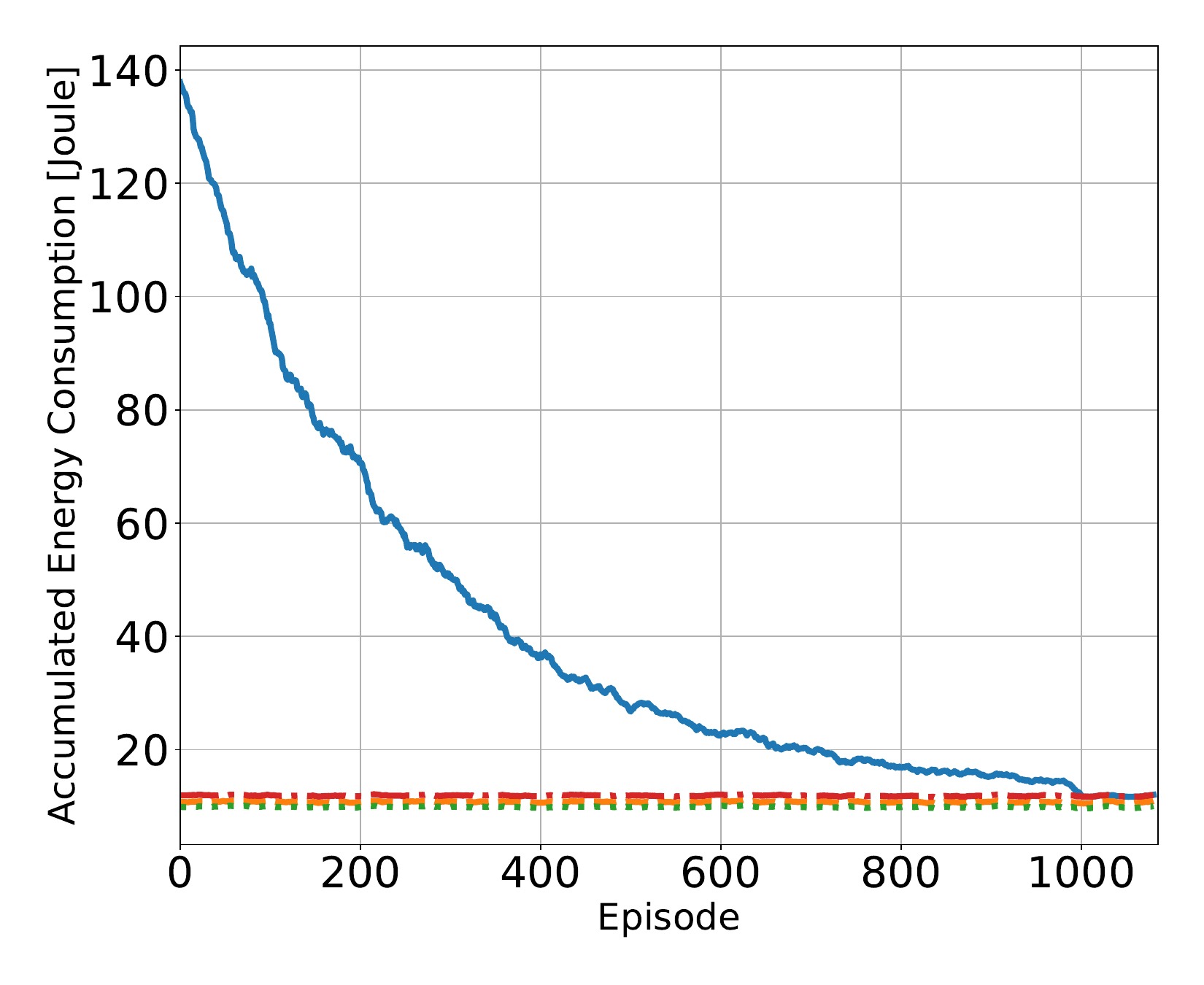}
        \caption{Accumulated Energy Consumption}
        \label{fig:mab-performance-analysis-b}
    \end{subfigure}
    \hfill
    \begin{subfigure}{.24\textwidth}
        \centering
        \includegraphics[width=\linewidth, trim= 0mm 20mm 0mm 10mm, clip=true]{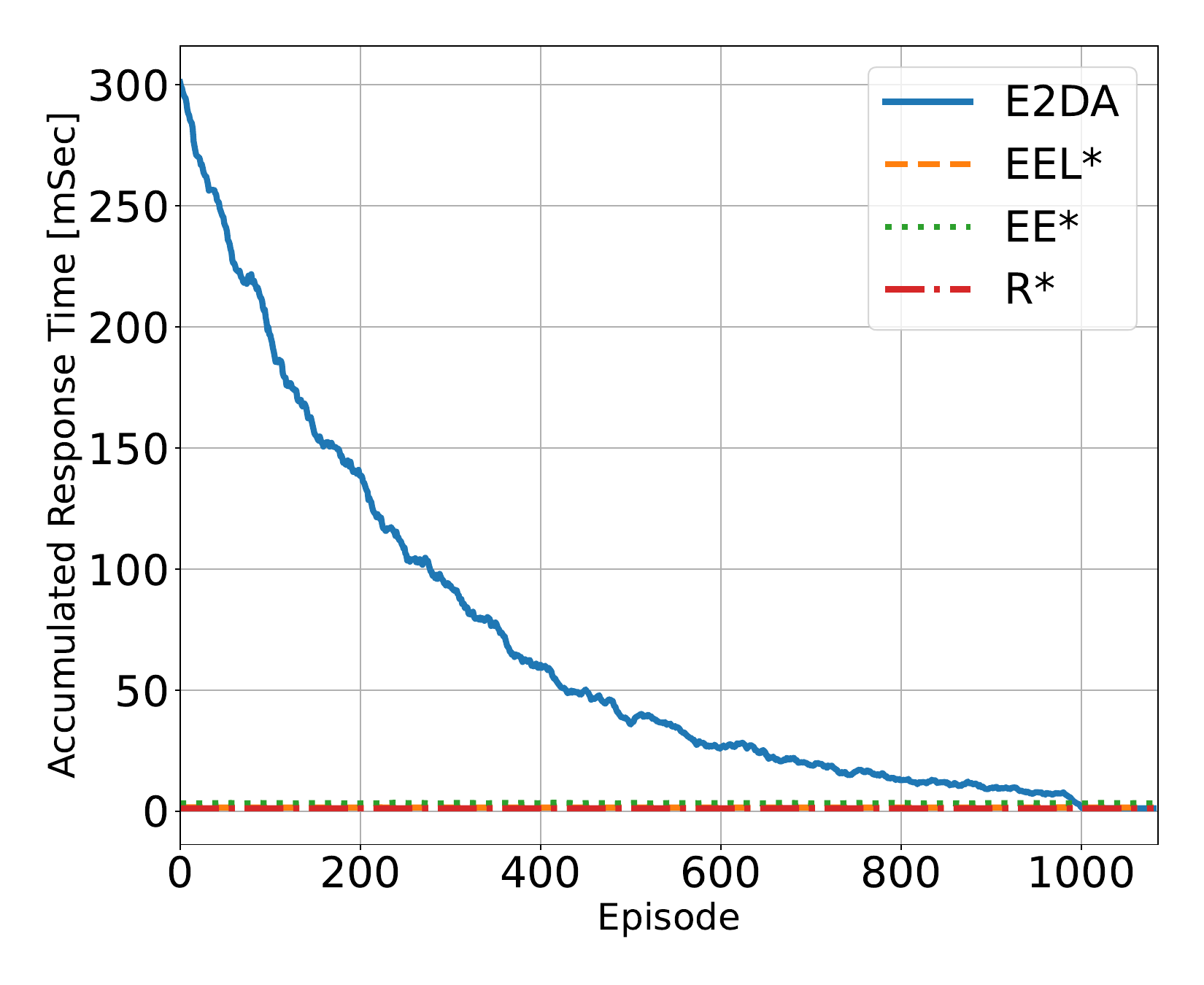}
        \caption{Accumulated Response Time}
        \label{fig:mab-performance-analysis-c}
    \end{subfigure}
    \caption{
    Performance of E2DA method in comparison with the baselines during the learning process
    in an environment with $\nbUAVs = 5$ edge users and $\nbMECservers = 3$ base stations. 
    A moving average with a window size of 20 has been applied to these plots to smooth out the fluctuations.}
    \label{fig:mab-performance-analysis}
    \vspace{-.2in}
\end{figure*}

\textbf{{Dataset Generation and Training of E2DA:}}
We used the developed simulator to generate a set of computation tasks for E2DA training.
For each task generated by the simulator, we measured waiting time, computation time, transmission time, response time, and energy consumption for both transmission and computation alongside task features (\ie task size, computational intensity, and deadline). We combined all these tasks to form a dataset containing a total of 32,565 tasks. 
Given the generated dataset, our E2DA agent is modeled 
by a two-layer neural network with 50 neurons on each layer. We use ReLu as activation function for hidden layers, and Sigmoid as activation function of the output layer.
The input is the task features, which is normalized between zero and one, and the output is the predicted reward obtained by performing each action, which is also normalized between zero and one.
We use the mean squared error as the loss function for this network and use the RMSprob optimizer to train the network for 1000 episodes and find the optimal parameters of the network. 
After the training is done, the E2DA agent is set to exploit (\ie testing phase) the learned policy for another 100 episodes to evaluate the performance of the learned offloading policy.
In each episode, the E2DA agent sees 100 tasks, randomly taken from the dataset, and makes an offloading decision based on the task features.

\textbf{{Baseline Algorithms:}}
We compare E2DA with three baselines, namely \emph{(1) Optimal Energy Efficient and Latency Baseline (\textit{EEL*})}, \emph{{(2) Optimal Energy Efficient Baseline (\textit{EE*})}}, and \emph{{(3) Optimal Response Time Baseline (\textit{R*})}}.
Each of these policies has access to a piece of unobservable information during the decision-making stage, which makes them superior to E2DA.
\textit{EEL*}
provides the optimal offloading strategy (\ie the action with the highest value of 
$\taskSize / (\taskResponse \totalEnergy^c)$) for the optimization problem (Eq. \ref{eq:optimization-problem}) without considering the deadline constraint. 
\textit{EE*}
makes an offloading decision by choosing the action that yields the highest $\taskSize / \totalEnergy^c$, and
\textit{R*} 
makes an offloading decision by choosing the action that yields the lowest task response time $\taskResponse$.


\subsection{Performance Evaluation}

\textbf{Training Performance:}
Fig. \ref{fig:mab-performance-analysis} shows the performance of  E2DA in comparison with the optimal baselines during the training and testing phases. 
To generate a dataset for this case, we set $\nbUAVs = 5$ edge users and $\nbMECservers = 3$ base stations.
Fig. \ref{fig:mab-accumulated-reward} shows the E2DA accumulated episode reward.
The E2DA, after 1,000 episodes of training, outperforms both \textit{R*} and \textit{EE*} and converges to \textit{EEL*}. 
Fig. \ref{fig:mab-performance-analysis-a} demonstrates the fraction of tasks that have met their deadlines in each episode of learning, which increases as the agent progresses toward the optimal policy.
Fig. \ref{fig:mab-performance-analysis-b} shows the energy consumption profile of the edge user during the learning process.
\textit{R*} provides the worst energy consumption, while E2DA 
provides a performance very close to \textit{EEL*}.
Fig. \ref{fig:mab-performance-analysis-c} shows the accumulated tasks' response time for each episode. As expected, \textit{R*} provides the optimal response time, however, E2DA
converges to \textit{R*} and \textit{EEL*} in terms of response time.

\textbf{Effect of Task Distribution:}
As mentioned earlier, emerging computationally intensive applications (\eg{AR/VR, autonomous cars, conversational robots}) are increasingly being deployed at the network edge.
Tasks generated by each application/user have different characteristics. 
For example, AR/VR devices generate tasks at a high rate with large sizes that require high computational power (larger computational intensity); A conversational robot, such as ChatGPT, however, generates tasks at a lower rate with small sizes (\ie a question in the form of a sentence asked by a user) that need higher computational power. 
In this section, we analyze the effect of each of the tasks features on energy consumption and response time.

\begin{figure}[!t]
    \centering
    \begin{subfigure}{.49\linewidth}
        \centering
        \includegraphics[width=\linewidth, trim= 0mm 20mm 0mm 10mm, clip=true]{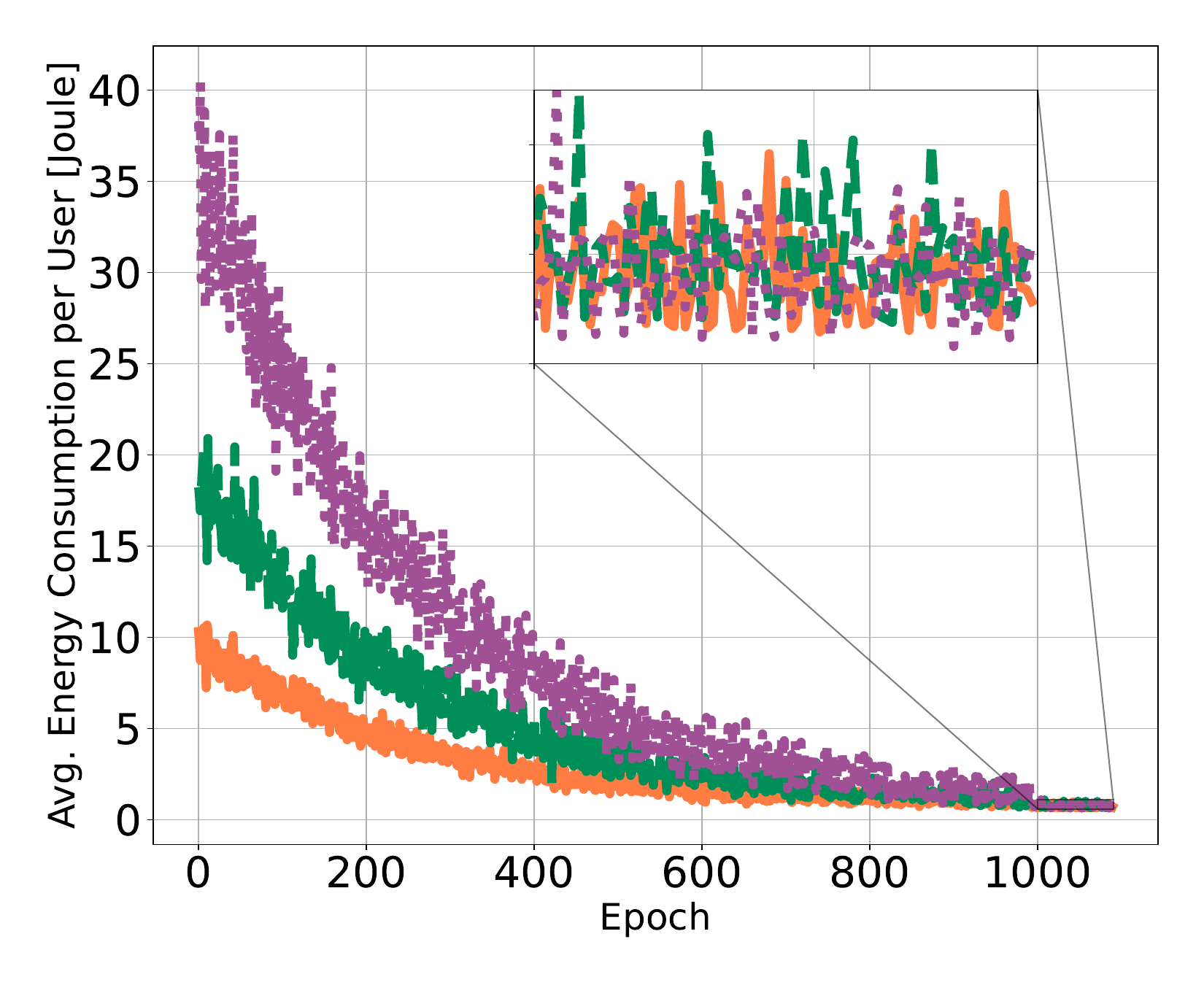}
        \caption{Avg. Energy Consumption}
        \label{fig:intensity-analysis-a}
    \end{subfigure}
    \hfill
    \begin{subfigure}{.49\linewidth}
        \centering
        \includegraphics[width=\linewidth, trim= 0mm 20mm 0mm 10mm, clip=true]{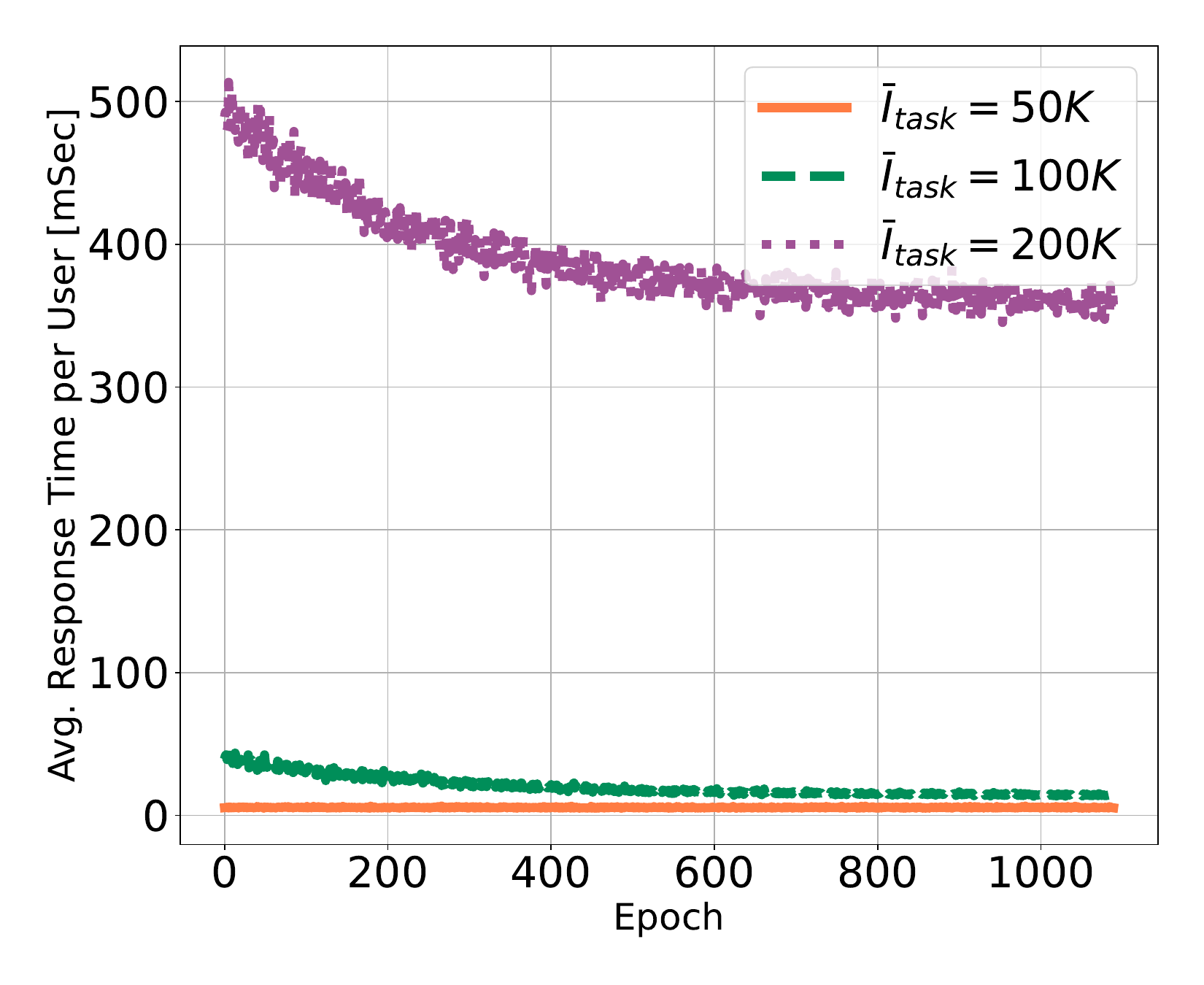}
        \caption{Avg. Response Time}
        \label{fig:intensity-analysis-b}
    \end{subfigure}
    \caption{Effect of task computational intensity.
    $\nbUAVs = 10$ edge users and $\nbMECservers = 3$ edge servers. In all cases, the mean task size is set to $\bar{S}_{task} = 5 \; K bit$ and the mean arrival rate is set to $\lambda = 20 \; task/sec$. 
    }
    \label{fig:intensity-analysis}
    \vspace{-.15in}
\end{figure}

Fig. \ref{fig:intensity-analysis-a} reveals that 
a 4X increase in computational intensity ($50$K vs. $200$K) results in an average of 1.04X increase in energy consumption during the testing phase.
This means that an increasing task computational intensity has little to no effect on the energy consumption.
Fig. \ref{fig:intensity-analysis-b}, on the other hand, shows that a 2X and 4X increase in computational intensity results in an average of 2.6X and 65.6X increase in response time, respectively.
This means that computational intensity has a significant impact on the response time due to the increased computation and queuing times on both edge and local processing units. 
In this analysis, the mean task size is set to $\bar{S}_{task} = 5Kbit$ and the mean arrival rate for all the ten users is set to $\lambda_k = 20 \; task/sec$. 

\begin{figure}[!h]
    \centering
    \begin{subfigure}{.49\linewidth}
        \centering
        \includegraphics[width=\linewidth, trim= 0mm 20mm 0mm 10mm, clip=true]{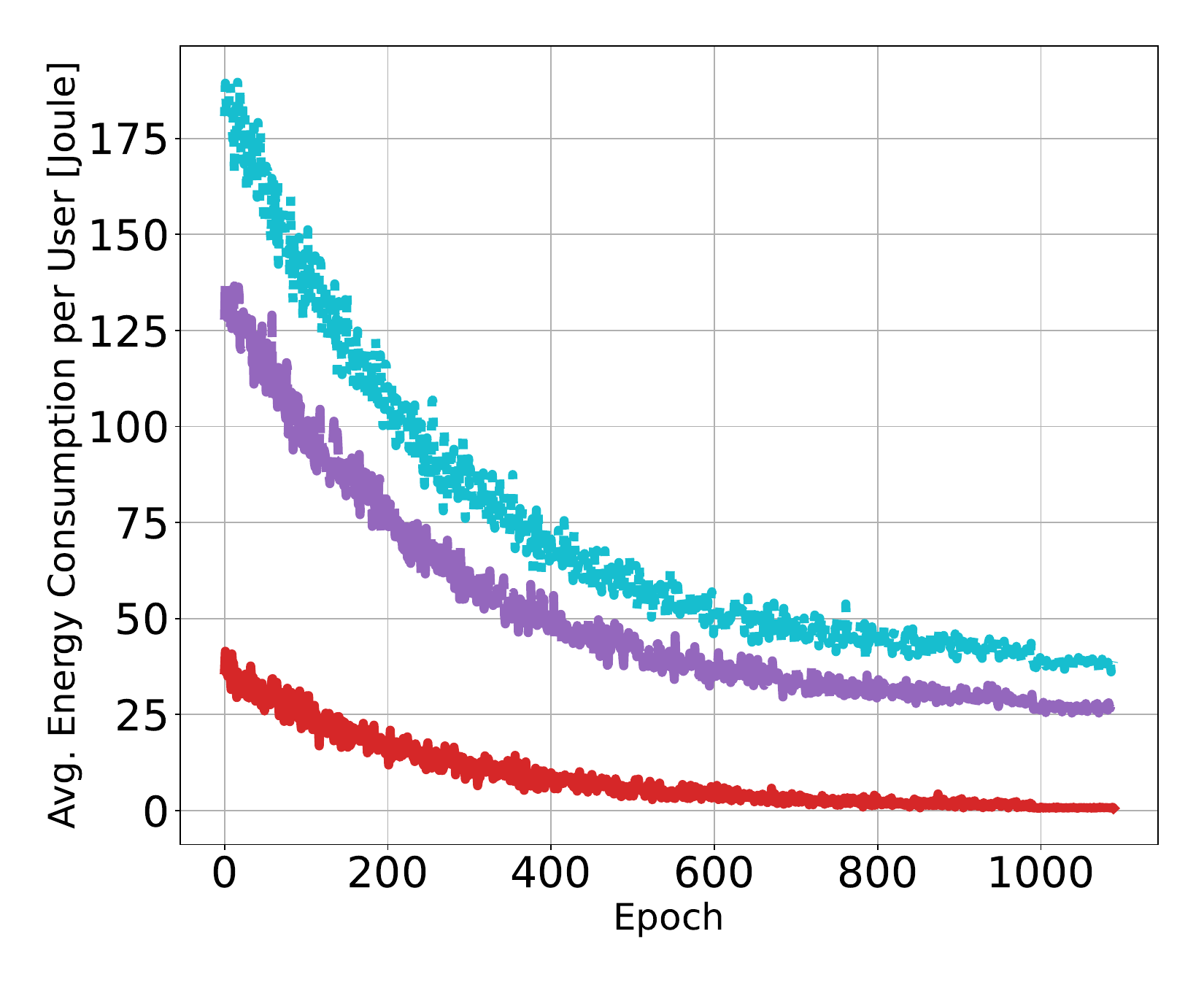}
        \caption{Avg. Energy Consumption}
        \label{fig:size-analysis-a}
    \end{subfigure}
    \hfill
    \begin{subfigure}{.49\linewidth}
        \centering
        \includegraphics[width=\linewidth, trim= 0mm 20mm 0mm 10mm, clip=true]{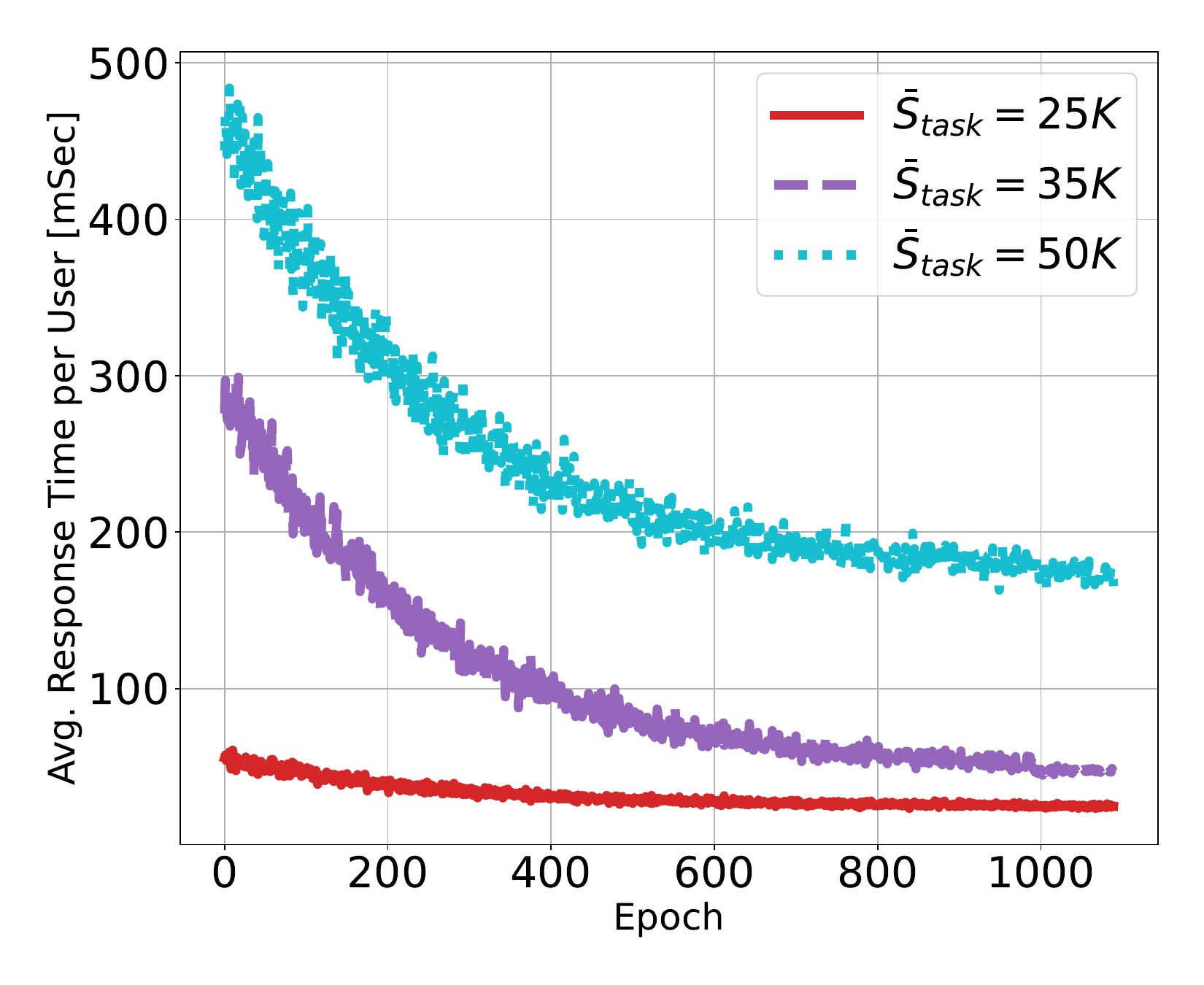}
        \caption{Avg. Response Time}
        \label{fig:size-analysis-b}
    \end{subfigure}
    \caption{Effect of task size. 
    $\nbUAVs = 10$ edge users and $\nbMECservers = 3$ edge servers. In all cases, the mean tasks' arrival rate is set to $\lambda = 60 \; task/sec$ and the mean computational intensity is set to $\taskIntencity = 5K\; cycle/bit$. 
    }
    \label{fig:size-analysis}
    \vspace{-.2in}
\end{figure}

Fig. \ref{fig:size-analysis} demonstrates the effect of increasing task size on the average task energy consumption and  task response time. The results show that increasing the task size by 2X leads to an almost 50X increase in energy consumption and 7X increase in response time.
This reveals that task size is a major contributing factor in increasing both energy consumption and response time. 
In this analysis, the mean tasks' arrival rate is set to $\lambda_k = 60 \; task/sec$ and the mean computational intensity is set to $\bar{I}_{task} = 5K\; cycle/bit$ (\eg AR/VR applications). 

\section{Conclusion}\label{sec:conclusion}
In this paper, we considered the problem of partially observable decentralized multi-channel task offloading, where $\nbUAVs$ edge users choose to offload their tasks to a nearby base station using one of the $\nbChannels$ wireless communication channels or compute the task locally ($C+1$ actions in total). The overall objective is to optimize energy efficiency while meeting each task deadline. However, the system is partially observable in the sense that not all the system states (channel rates, network traffic, etc.) are available at the decision-making stage.
Leveraging the CMAB models, we developed the E2DA algorithm that utilizes task features (\ie task size, computational intensity, and deadline) to make an offloading decision using one of the $C$ available channels, or perform the computation locally, without a need for full state information. 
Through numerical results we showed that the E2DA policy learns the offloading policy that converges to optimal baseline solutions. 
Moreover, we demonstrated that task computational intensity has little impact on edge users’ energy consumption and a larger impact on response time, while task size has a significant impact on both energy consumption and task response time due to an increased transmission and computation time.

\vspace{-.1cm}
\section{Acknowledgement}
\vspace{-.1cm}
The material is based upon a work supported by the National Aeronautics and Space Administration (NASA) under award No(s) 80NSSC20M0261, and the National Science Foundation (NSF) grants CNS-1948511 and CNS-1955561. 
Any opinions, findings, and conclusions or recommendations expressed in this material are those of the author(s) and do not necessarily reflect the views of NASA and NSF. 

{\small
\bibliographystyle{IEEEtranN}
\bibliography{ref2}
}
\end{document}

%% file: notation.tex
\newcommand\task{T}
\newcommand\taskArrival{\task_{a}}
\newcommand\taskTxTime{\task_{tx}}

\newcommand\taskWaitTime{W}

\newcommand\taskExecution{\task_{e}}

\newcommand\taskSize{S_{task}} 
\newcommand\taskIntencity{I_{task}} 
\newcommand\taskDeadline{\task_{d}} 
\newcommand\taskResponse{\task_{r}} 

\newcommand\resultSize{S_{res}} 
\newcommand\resultTxTime{\task_{rx}}

\newcommand\nbUAVs{K}
\newcommand\nbChannels{C}
\newcommand\nbMECservers{N}

\newcommand\cpuFreq{f} 
\newcommand\mecCPUFreq{\cpuFreq_{mec}} 
\newcommand\uavCPUFreq{\cpuFreq_{eu}} 

\newcommand\channelRate{R}
\newcommand\ulChannelRate{\channelRate_{k, n}}
\newcommand\dlChannelRate{\channelRate_{n, k}}

\newcommand\power{P}
\newcommand\ulPower{\power_{k, n}}
\newcommand\dlPower{\power_{k}}

\newcommand\energy{E}
\newcommand\totalEnergy{\energy_{tot}}
\newcommand\cpuEnergy{\energy_{e}}
\newcommand\txEnergy{\energy_{tx}}
\newcommand\rxEnergy{\energy_{rx}}

\newcommand{\expected}[1]{{\mathop{\mathbb{E}} \left[#1\right]}}

\newcommand\indicator{\mathbbm{1}}

\newcommand{\ie}{{\textit{i.e., }}}
\newcommand{\eg}{{\textit{e.g., }}}